# PRECISE POINTING OF CUBESAT TELESCOPES: COMPARISON BETWEEN HEAT AND LIGHT INDUCED ATTITUDE CONTROL METHODS

**Ravi teja Nallapu[*] and Jekan Thangavelautham[†]**

CubeSats are emerging as low-cost tools to perform astronomy, exoplanet searches and earth observation. These satellites can target an object for science observation for weeks on end. This is typically not possible on larger missions where usage time is shared. The problem of designing an attitude control system for CubeSat telescopes is very challenging because current choice of actuators such as reaction-wheels and magnetorquers can induce jitter on the spacecraft due to moving mechanical parts and due to external disturbances. These telescopes may contain cryo-pumps and servos that introduce additional vibrations. A better solution is required. In our paper, we analyze the feasibility of utilizing solar radiation pressure (SRP) and radiometric force to achieve precise attitude control. Our studies show radiometric actuators to be a viable method to achieve precise pointing. The device uses 8 thin vanes of different temperatures placed in a near-vacuum chamber. These chambers contain trace quantities of lightweight, inert gasses like argon. The temperature gradient across the vanes causes the gas molecules to strike the vanes differently and thus inducing a force. By controlling these forces, it's possible to produce a torque to precisely point or spin a spacecraft. We present a conceptual design of a CubeSat that is equipped with these actuators. We then analyze the potential slew maneuver and slew rates possible with these actuators by simulating their performance. Our analytical and simulation results point towards a promising pathway for laboratory testing of this technology and demonstration of this technology in space.

**INTRODUCTION**

A spacecraft's attitude is typically controlled by reaction-wheels[1], which are flywheels that trade their angular momentum with that of the spacecraft. However, as the reaction-wheels fight disturbance torques, they keep accumulating angular momentum; and eventually they reach their maximum rated angular momentum reaching saturation[2]. Despite low disturbance torques in space, the reaction wheel saturation is a problem for extended space missions in deep space such space telescope missions. Techniques to desaturate a reaction-wheel do exist such as magneto-torquer desaturation using the Earth's magnetic field or propulsion based desaturation in deep space[3]. However, in the absence of a planetary magnetic field[4], the magnetorquers become non-functional. Other techniques such as using an additional reaction wheel[5] becomes both expensive and limit the allowable mass and volume of the spacecraft. Hence there is a need for an attitude control actuator that is functional in deep-space and does not require propellant.

---

[*] PhD Student, Space and Terrestrial Robotic Exploration Laboratory, Arizona State University, 781 E. Terrace Mall, Tempe, AZ.
[†] Assistant Professor, Space and Terrestrial Robotic Exploration Laboratory, Arizona State University, 781 E. Terrace Mall, Tempe, AZ

For an Earth orbiting spacecraft, typical disturbances include atmospheric drag[6], gravity gradient[7] and solar radiation pressure (SRP)[8]. However, in deep space the atmospheric drag and gravity gradient can be eliminated, leaving the SRP as the only known dominant disturbance. The SRP is the pressure exerted by the solar flux on an object in space and depends on the optical properties of the object exposed to the photons, such as specular reflectivity and absorptivity. Hence, one strategy to control the attitude can be to manipulate these properties, thus controlling the magnitude of the disturbance.

The IKAROS spacecraft[9], launched by JAXA in 2010 used this technique to maneuver itself to Venus. Their spacecraft had a large, square solar-sail of diagonal length 20 m, and LCD cells embedded along the perimeter of the sail as shown in Figure 1(left) to control the attitude of the spacecraft. These LCD cells are turned "on" and "off", which varies the reflectivity, thus changing the SRP acting upon the surface.

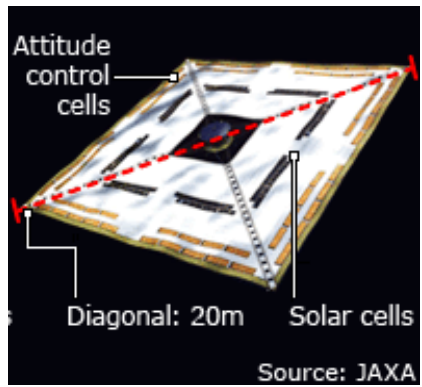 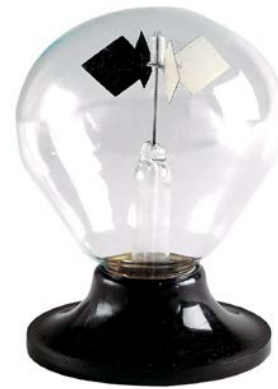

**Figure 1. Technologies that exemplify photometric and radiometric forces: IKAROS Spacecraft (left) and Crookes Radiometer (Right)**

In retrospect, the binary operation of switching the LCDs "on" or "off" can be further improved by advances in technology. Devices like the Transflector®[10] and Smart Windows®[11] that switch states from a transparent glass to a mirror present interesting alternatives.

Another technology which has not yet been applied to space missions, but can be of great importance is radiometric forces, which are exemplified by the Crookes Radiometer[12], shown in Figure 1(right). The Crookes Radiometer consists of 4 plates, called vanes that are colored white and black on opposite ends. The vanes are mounted on a spindle, and placed inside a partial vacuum chamber, typically consisting of trace amounts of argon. When the setup is exposed to sunlight, the vane-spindle starts to spin. This phenomenon has been widely misunderstood, with many proposing photons hitting the vanes and causing the force. However it is Thermal Creep force[13], which is a radiometric force that causes the vane to spin and is further described in the paper.

This paper intends to apply and compare the performance of these two phenomena: light and heat to the small satellite attitude control problem. Specifically, we compare the magnitudes of these two forces, in their scalar form, and then analyze their application to an on-orbit centrifuge science laboratory mission such as AOSAT 1[14,24,25] using a CubeSat[17]. We begin by briefly introducing the two phenomena. Following this, the application of this technology to CubeSats is discussed by designing hypothetical actuators. We then compare the magnitudes of the forces fol-lowed by performance comparison of these two actuators in the results section. Finally, we summarize the findings in the conclusion section.

## PHYSICS OF THE PHENOMENA

### Solar Radiation Pressure Force

The SRP force is caused by photons striking the surface of a spacecraft. Consider a spacecraft whose position vector relative to the sun is $\vec{s}$ as shown in Figure 2. Let $A$ be the area of each face, and let the face under consideration have an outward normal vector whose direction is $\hat{n}$. Let the normal subtend an angle $\theta$ with the unit vector $\hat{s}$ (direction of $\vec{s}$).

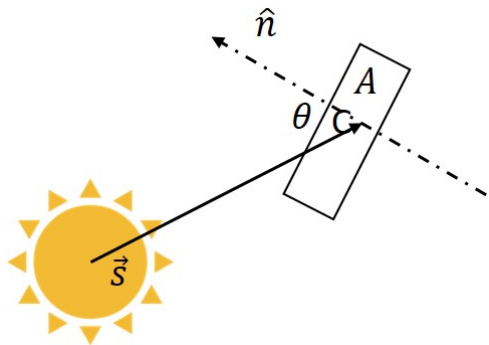

**Figure 2. CubeSat face exposed to solar radiation.**

In this case, the magnitude of SRP[15] acting on the whole spacecraft is given by:

$$\vec{F}_{SRP} = \frac{G_1}{|\vec{s}|^2} \sum_i A_i \cos\theta_i \left[ (1 - c_{sr})\hat{s} + 2\left(\frac{c_{dif}}{3} + c_{sr}\cos\theta_i\right)\hat{n}_i \right] \quad (1)$$

Here $G_1$ denotes the solar constant whose value is 1E14 kg-km/s$^2$, and '$i$' denotes the index of all those faces that face the sun, i.e., $\hat{s}\cdot\hat{n}_i > 0$. Since the SRP depends on the inclination between the sun-vector and the spacecraft normal, we look at the maximum value of SRP, which occurs when $\theta_i = 0$ Degrees, i.e., when $\hat{s}$ aligns with $\hat{n}$. Hence the maximum available SRP is given by:

$$\vec{F}_{Max\,SRP} = \frac{G_1}{|\vec{s}|^2} \sum_i A_i \left[1 + \frac{2}{3}c_{dif} + c_{sr}\right]\hat{n}_i \quad (2)$$

The positive coefficients, $c_{dif}$ and $c_{sr}$, when added constitute the total coefficient of reflection which varies between 0 and 1 and hence we can relate these coefficients as:

$$c_{dif} + c_{sr} \leq 1 \quad (3)$$

### Radiometric Forces

The radiometric force is observed when a gas is exposed to a surface with a temperature gradient. In the above example, the gradient is created by the vanes as the black side of the vane absorbs more light and becomes hotter than the white side. In this setup, two forces are generated[13]: A normal pressure force exerted by the gas molecules as they strike the black side with higher momentum, and a thermal creep force occurs as the gas molecules strike the edges of the vanes. The two forces are shown in Figure 3. The net radiometric force then is given by:

$$F_R = F_N + F'_S \qquad (4)$$

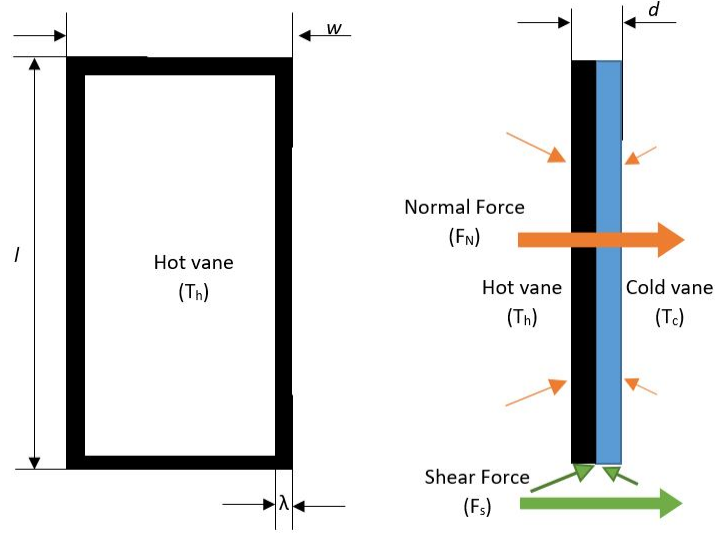

**Figure 3. Radiometric forces acting on a vane with a temperature gradient**

Consider a vane with a temperature gradient as shown in Figure 3, where the hotter side is at a temperature $T_h$ and the colder side is at $T_c$. Let $l$, $w$, and $d$, represent the length, width, and thickness respectively. Let this vane be placed in a rarefied gas with a gas kinetic diameter $\sigma$. In this case, the normal force[16] is given by:

$$F_N = (2-\alpha)\frac{15}{32\sqrt{2}\pi}\frac{k}{\sigma^2}\frac{\Delta T}{d}(\lambda P) \qquad (5)$$

where, $\alpha$ is the coefficient of thermal accommodation, $\Delta T$ is the temperature difference $T_h - T_c$, $P$ is the perimeter of the vane, and $k$ is the Boltzmann constant. The Thermal Creep force[17] is given by:

$$F'_s = \frac{x15}{32\sqrt{2}\pi}\frac{k\Delta T\alpha}{\sigma^2}(l' + w')\min\left(\frac{d}{w_{grad}}, 1\right) \qquad (6)$$

where, $x$ is a correction factor to compensate for slip reduction due to opposing forces, $l'$ and $w'$ are reduced length and width of the vane due to non-reactive drag forces given by equations 7 and 8 respectively and are associated with a numerical factor $\beta$. Also, $w_{grad}$ is the width of the gradient given by equation 9. The thermal creep force only comes into play when the thickness of the vane is comparable to the gas mean free path $\lambda$.

$$l' = l - \beta\lambda\left(\frac{2-\alpha}{\alpha}\right) \qquad (7)$$

$$w' = w - \beta\lambda\left(\frac{2-\alpha}{\alpha}\right) \qquad (8)$$

$$w_{grad} = d + 2\lambda\left(\frac{2-\alpha}{\alpha}\right) \qquad (9)$$

Both the pressure force (equation 5) and the creep force (equation 6) are directed from the hot side towards the cold side. We refer the readers to Scandurra[16] and Wolfe[17] for a complete derivation of equations 5 through 9.

**Application to Spacecrafts**

In this section, we describe the actuators that use light and heat respectively to control the attitude of a 3U CubeSat[18].

**SRP based actuators.**

We let the chassis walls of the spacecraft be covered by a switchable glass-mirror technology such as an E-TransFlector®, and the spacecraft walls are embedded with sun sensors[19] that can track the sun vector. With this setup, a feedback controller can be developed to orient the spacecraft as desired. A conceptual design of this spacecraft is shown in Figure 4.

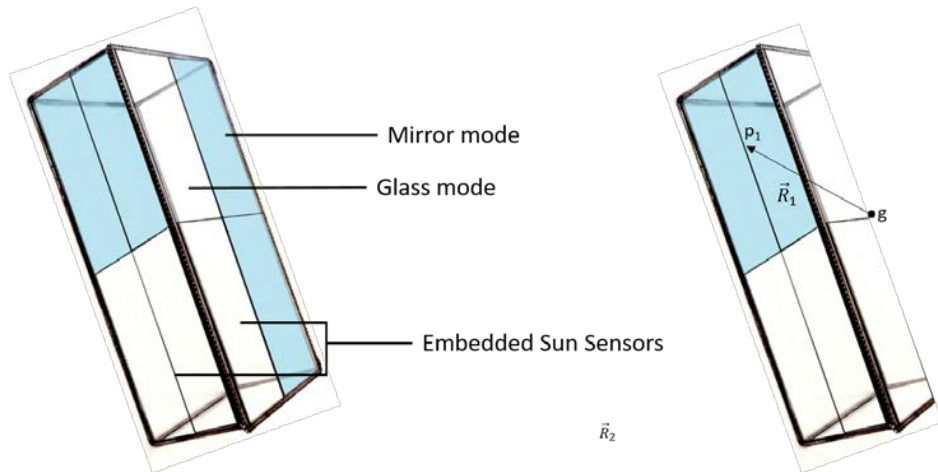

**Figure 4. Conceptual spacecraft utilizing solar radiation pressure to control its attitude (left). A cross-section view shows the moment arm (right).**

In Figure 4, the blue tinted areas represent the surface that has transformed to a mirror, while the white surfaces are transparent. By switching between the 'mirror' and 'transparent glass' modes we can induce SRP as desired. The moment arm vector, $R_1$, joins the center of gravity $g$ to the center of pressure $p_1$.

**Radiometric Actuators**

Let's consider an alternate attitude actuator that exploits the radiometric force[20]. We let the chassis walls of the CubeSat be covered with a glass chamber filled with trace amounts of argon. The chamber contains 8 vane plates arranged in a 2×2 matrix. The 2 matrices touch each other, and are affixed to the glass chamber as shown in Figure 5. The vanes are equipped with Thermo-Electric Devices (TED) such as the one mentioned in Reference 21 and temperature sensors so that their temperature can be controlled and measured.

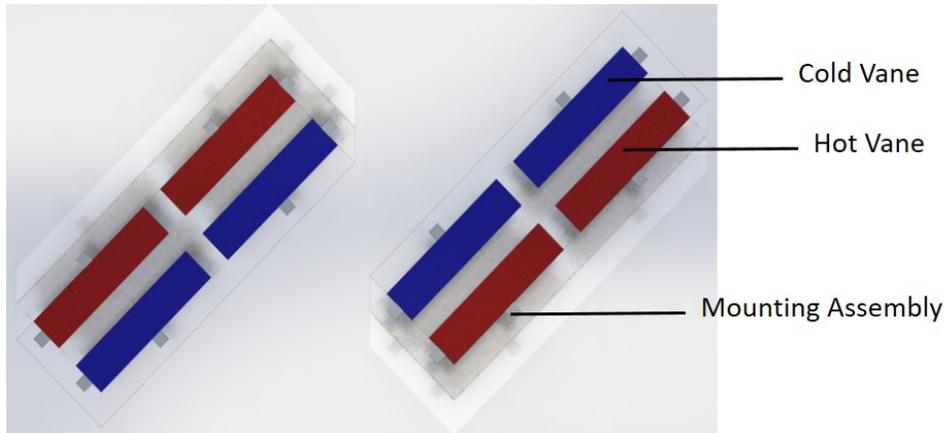

**Figure 5. Conceptual Radiometric actuator, Front view (left) and back view (right).**

Figure 6 shows the actuators fixed onto the walls of a 3U CubeSat, along with the moment arm vector shown on the right.

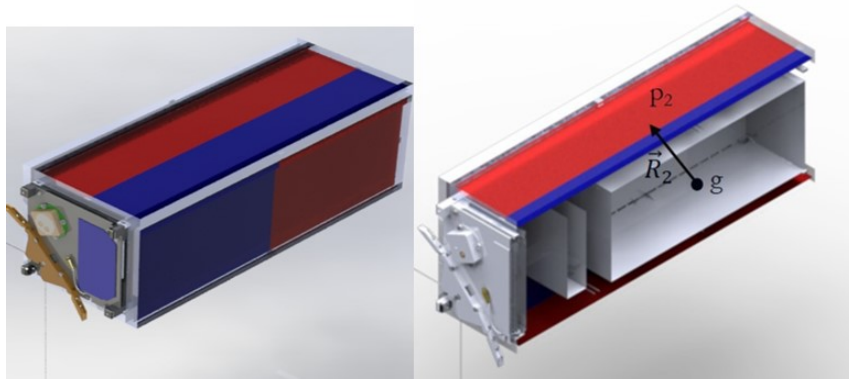

**Figure 6. Conceptual CubeSat with radiometric actuators (left) with an cross-section view to show the moment arm (right).**

Vanes in red represent the 'hot' vanes while vanes in blue represent the 'cold' vanes. Also, as seen in Figure 3, the vanes in contact have contrasting temperature to create the temperature gradient. The center of mass of the spacecraft is shown as $g$, while $p_2$ is the center of pressure on the face shown. The moment arm vector $R_2$ joins $g$ and $p_2$.

## PERFORMANCE COMPARISION

### Force Magnitudes

Consider a single face of the SRP actuator, and 1 hot-cold vane pair shown in Figures 5 and 6. Let the two faces being compared have the same geometry. The parameters considered for this simulation are presented in Table 1.

**Table 1. Simulation Parameters**

| Parameter | Value (Units) |
|---|---|
| Length | 30 cm |
| Width (vane-pair and SRP film) | 10 cm |
| Each length | 15 cm |
| Thickness (vane-pair and SRP film) | 0.1 cm |
| Distance from Sun | 1 AU |
| Gas for Radiometric actuator | Argon |
| Thermal accommodation coefficient | 0.83 |
| x | 0.5 |
| b | 0.5 |

The magnitude of the SRP force and radiometric forces are shown in Figure 7.

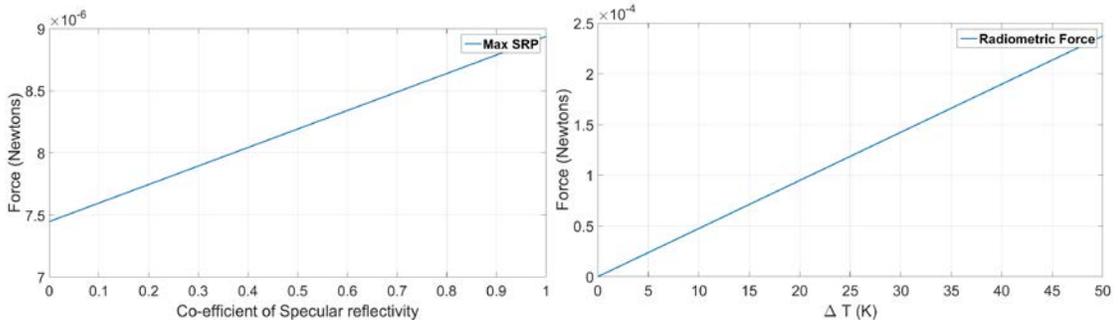

**Figure 7. Comparing magnitudes of maximum SRP (left) and radiometric forces (right)**

As seen here, the radiometric force can be up to 2 orders of magnitude greater than the maximum available SRP. It should be noted that the maximum SRP cannot be realized because of non-alignment with the sun vector and due to the inequality of equation 3. On the contrary, the radiometric forces can vary linearly with temperature as shown in Figure 7 (right).

**Single Axis Spin**

Consider the same single faces of actuators shown in Figures 5 and 6, now let the SRP face be modified such that the top half is purely specular ($c_{sr}=1$), and its bottom half is purely absorptive ($c_{sr}=0$), this is to ensure that the SRP torque generated is a maximum. Also, let the radiometric actuator being compared have 4 vanes that are arranged as 2 2×1 matrices facing each other, creating a temperature gradient between them. In the case of the radiometric actuator we assume a temperature gradient of $\Delta T = 5^{\circ}C$ across the thickness and perform a dynamics simulation in software. The two actuators are shown in Figure 8, along with the direction of the intended slew.

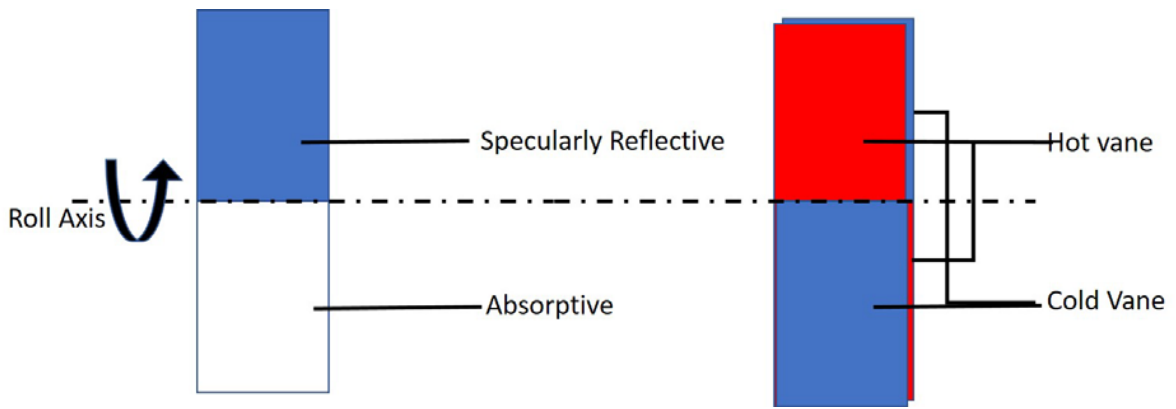

**Figure 8. Two actuators being compared: SRP (left) and Radiometric (Right)**

Attitude propagation was done by solving Euler's equations[22] with quaternions[23]. In this case, the spin is about the body *x*-axis as shown in Figure 8, which is the spacecraft roll. The propagated roll angles in both the cases are presented in Figure 9.

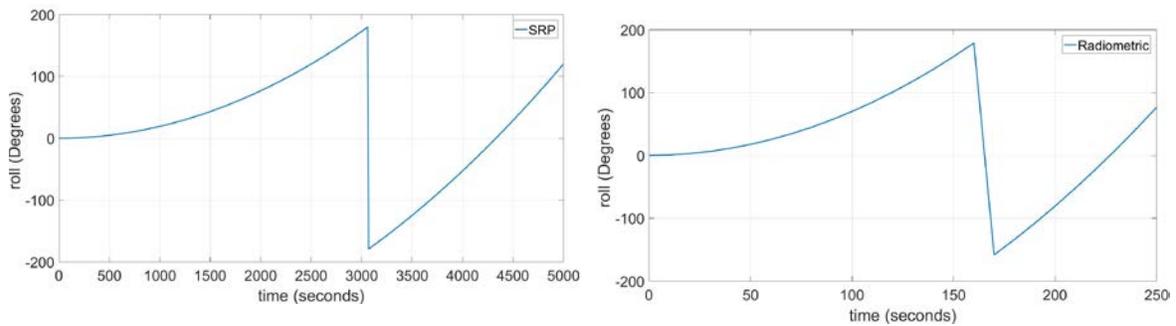

**Figure 9. Roll propagation with SRP (left) and radiometric forces (right)**

As seen here, it takes about 72 minutes to perform a complete 360° rotation with SRP, whereas, it only takes about 3.5 minutes with a radiometric force actuator which is a 20-fold improvement.

Most spacecraft missions can just be realized with slew maneuvers, however with science-led missions like AOSAT 1, there is a critical need to perform attitude maneuvers throughout the mission. AOSAT 1 for instance will be spinning at 1 RPM in Low Earth Orbit (LEO). We examine the applicability of these methods to AOSAT 1. Specifically, we compare the angular velocity response of the two methods (Figure 10).

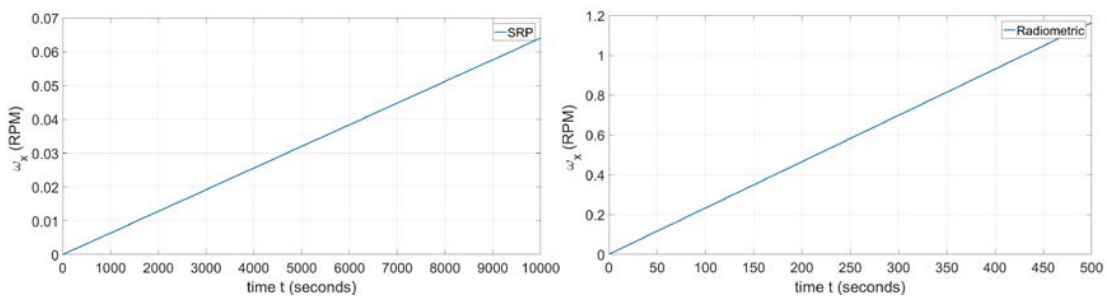

**Figure 10. Angular velocity propagation with SRP (left) and radiometric forces (right)**

Inferring from the slope of Figure 10 (left), it takes an SRP based system about 2 days to reach up to 1 RPM, which makes it impractical. On the other hand, a radiometric actuator system can reach 1 RPM within 7 minutes with a $\Delta T$ of 5$^o$C, as seen in Figure 10 (right). This shows that the radiometric actuator has a 400-fold advantage over SRP.

**CONCLUSION**

This paper compares two new methods to control attitude of a spacecraft, using solar radiation pressure method (light) and radiometric method (heat), for deep space missions at 1 AU. The magnitudes of the maximum available solar radiation pressure force and the radiometric forces were compared. It was found that the radiometric actuator for a temperature gradient of about 50$^o$C, can provide a force that is 2 orders of magnitude greater than the maximum solar radiation pressure force. The two actuators were also compared based on their roll and slew rate performance. It was found that the SRP actuators were much slower when compared to the radiometric actuators. These results lead us to the conclusions that the solar radiation pressure actuators are slow, and that can be applied to missions that need precise pointing, because of the finer slew rates. However, due to the inverse square law dependence of the solar distance, their application can be practical only in the inner solar system. On the other hand, the radiometric actuators provide greater force magnitudes, and therefore faster slew and spin performance. Also, since these are not dependent on an external source, the radiometric actuators can work anywhere in deep space. However, the challenge of creating the required temperature gradient does exist, and will be fully addressed in a future paper.